\documentclass[aps,prb,showpacs,twocolumn,superscriptaddress,amsmath,amssymb]{revtex4-1}

\usepackage{graphicx}
\usepackage{bm}
\usepackage{xspace} 
\usepackage{dsfont} 
\usepackage{psfrag}
\usepackage{bbm}
\usepackage{hyperref}  
\usepackage{float}

\usepackage[usenames]{color}
\usepackage{textcomp}
\newcommand{\ket}[1]{|#1\rangle}

\newcommand{\unity}{{\mathbbm{1}}}

\begin{document}

\title{Signatures of Delocalization in the Fermionic 1D Hubbard Model with Box Disorder: \\ Comparative Study with DMRG and R-DMFT}
\author{Julia Wernsdorfer}
\affiliation{Institut f\"ur Theoretische Physik, Johann Wolfgang Goethe-Universit\"at, 60438 Frankfurt/Main, Germany}
\author{Georg Harder}
\affiliation{Institut f\"ur Theoretische Physik C, RWTH Aachen University, 52056 Aachen, Germany}
\author{Ulrich Schollw\"ock}
\affiliation{Physics Department, Center for NanoScience and Arnold Sommerfeld Center for Theoretical Physics, Ludwig-Maximilians-Universit\"at M\"unchen, D-80333 M\"unchen, Germany}
\author{Walter Hofstetter}
\affiliation{Institut f\"ur Theoretische Physik, Johann Wolfgang Goethe-Universit\"at, 60438 Frankfurt/Main, Germany}
\date{\today}

\begin{abstract}
We investigate the 1D Anderson-Hubbard model at half filling with box-disorder.
The ground state phase diagram is obtained by means of real-space dynamical mean-field theory (R-DMFT) and the density matrix renormalization group (DMRG). We find Mott insulating and Anderson localized regimes as well as a strong indication of a delocalized phase for intermediate interaction and disorder strength within accessible system sizes. These phases are characterized and distinguished by qualitatively different scaling behavior of the local density of states, the energy gap in the excitation spectrum and the inverse participation number. 
\end{abstract}


\maketitle
\section{Introduction}

Disorder is omnipresent in solids due to defects and impurities. In his breakthrough theoretical investigation of non-interacting disordered systems Anderson showed that coherent backscattering processes due to impurities are responsible for spatial localization of electronic wave functions.\cite{Anderson58} The scattering probability increases with the impurity density and beyond a critical value leads to  spatial localization of the electrons. If the states at the Fermi level become localized, the system becomes an Anderson insulator. Whereas in 3D systems the disorder strength has to overcome a certain threshold to drive the system from a metal to an insulator, in 2D and 1D systems any nonzero disorder strength is sufficient to cause localization in the thermodynamic limit.\cite{Abrahams, Lieb03}

On the other hand, a Mott-Hubbard metal-insulator transition in a pure system can be induced by electron-electron interaction for integer filling.\cite{Mott37, Imada98} In contrast to scattering by impurities, this many-body effect induces correlations and tends to distribute particles uniformly. Numerical and analytical investigations via dynamical mean-field theory, exact diagonalization, quantum Monte Carlo and analytical calculations revealed a critical interaction strength in 3D and 2D at which the phase transition takes place.\cite{Georges96,Mancini00,Kotliar04,michielsen} In one spatial dimension in the thermodynamic limit the Mott phase is present at any interaction strength.\cite{Lieb68, Lieb03, Giam_Mott}  

The physics resulting from the simultaneous presence of both effects is not yet fully understood. Although interaction and disorder both lead to metal-insulator transitions, their action is competing: the repulsive electron-electron interaction favors uniform distribution of the particles, while disorder localizes the electronic wave function to a few lattice sites. 

Several investigations have been performed on the competition between interaction and disorder in the Anderson-Hubbard model: such as a perturbative renormalization group (RG)\cite{Giamarchi88} analysis and non-perturbative treatments via density matrix renormalization group (DMRG) and dynamical mean-field theory (DMFT) techniques, to name just a few.\cite{Finkelstein84, Pang93, Byczuk05, Schwab09, Aguiar09, YunSong, semmler, speckle10} In these studies the interplay was found to affect the system in a subtle way. For a semi-elliptic density of states in high dimensions the presence of both disorder and interaction was found to lead to a disordered metal surrounded by an insulating phase, the latter consisting of Anderson and Mott insulators continuously connected with each other.\cite{Byczuk05, Aguiar09} Since in the first studies of non-interacting, weakly interacting, as well as strongly interacting 2D systems\cite{Abrahams,altshuler80,tanatar89} only insulating regions were found, it was believed for a long time that a metallic phase cannot occur. However, theoretical studies via RG  \cite{Finkelstein84,Punnoose} involving $N$ flavors of electrons predicted the existence of a quantum critical point, at which a metal-insulator transition takes place in 2D. This prediction was confirmed experimentally a few years later.\cite{anissimova} In 1D bosonic systems a Mott insulator, a Bose glass and a superfluid phase were found,\cite{Fisher89, Prokof04, Scalettar91, Prokofev-comment98, Rapsch99} of which the latter is analogous to the metallic phase in fermionic systems. However, for repulsively interacting fermions in 1D perturbative RG calculations predicted a random antiferromagnet, where the fermions localize individually around randomly distributed sites.\cite{Giamarchi88} Nevertheless, perturbative treatments of disorder can not be expected to capture the physics in the full range of couplings. In 1D up to now only next-neighbor interacting disordered spinless fermions have been studied exactly via DMRG\cite{Schmitteckert98} where, based on the phase sensitivity, a delocalized phase for  intermediate attractive interaction and disorder strength was found. The ground state phase diagram of spinful fermions with on-site interactions has not been determined yet.

The careful experimental analysis of these phenomena requires tunability of the disorder and interaction strengths. During the last decade it has become possible to simulate theoretical models of solid state systems by ultracold atoms in optical lattices.\cite{Greiner02, Schneider08, Joerdens08}  Due to the precise control over the system parameters and the possibility to artificially introduce different kinds of disorder, e.g. binary\cite{Esslinger06, Ospelkaus06}, bichromatic\cite{Damski03, Fallani07, Roux08} and speckle\cite{Billy08, White_speck}, these systems are highly suited for detailed investigations of disorder phenomena. In particular, the spatial dimensionality can be easily adjusted and the localization of bosonic matter waves in 1D\cite{Billy08, Roati08} and of fermions in 3D\cite{deMarco} has already been observed.

A non-perturbative investigation of a spin-$\frac{1}{2}$ 1D fermionic system with box disorder is the aim of this paper. By varying the disorder and  interaction strengths, Anderson- and Mott-insulating regimes as well as delocalization are found within our simulated system sizes,
based on the real-space extension of DMFT (R-DMFT) and on DMRG. For their characterization, the following physical observables were calculated: a) the geometric average of the local density of states, representing its typical value, and its scaling behavior with the system size, b) the charge gap in the thermodynamic limit, c) the inverse participation ratio and its dependence on the system size.  

The paper is organized as follows: in Sec.~\ref{sec:model} we introduce the Anderson-Hubbard Hamiltonian. The complementary DMRG and R-DMFT methods applied in this paper are briefly explained in Sec.~\ref{sec:method}. The resulting physical observables and a detailed analysis of the emerging delocalized, Anderson and Mott-Hubbard regimes are discussed in Sec.~\ref{sec:results}. In Sec.~\ref{sec:summary} we conclude with a summary of our results. 

\section{Model} \label{sec:model}	
\subsection{Hamiltonian}
Strongly correlated disordered fermions on a lattice are described by the Anderson-Hubbard Hamiltonian 
\begin{eqnarray}
 \mathcal{H}&=&-t\sum_{i,\sigma}(c_{i,\sigma}^\dag c_{i+1, \sigma} + c^\dag_{i+1,\sigma}c_{i,\sigma}) + U\sum_{i} n_{i\uparrow}n_{i\downarrow}\nonumber\\
 &&+ \sum_{i,\sigma} (\varepsilon_i-\mu) n_{i\sigma}\,,
\label{eq:Ham}
\end{eqnarray}
where $\sigma\in\{\uparrow,\downarrow\}$ labels spin, while $c^\dag_{i\sigma}$, $c_{i\sigma}$ and $n_{i,\sigma}$ are the creation, annihilation and particle number operators for an electron on site $i$ with spin $\sigma$. $U$ is the on-site interaction and $t$ is the nearest-neighbor hopping matrix element. The on-site energies $\epsilon_i$ are random variables, each distributed independently according to $\mathcal{P}(\varepsilon_i)=\Theta(D-|\varepsilon_i|)\cdot2/D$. Here $\Theta$ is a Heaviside function and $D$ is the disorder strength. We consider a 1D bipartite lattice with commensurate filling $\langle n_i \rangle=\langle n_{i\uparrow}+n_{i\downarrow}\rangle=1$. 
In the homogeneous case, i.e. for $D=0$, this Hamiltonian can be solved exactly by means of the Bethe ansatz.\cite{Lieb68} However, advanced numerical methods are required when the on-site energies are random. 

\section{methods}\label{sec:method}

Commonly used approaches to solve interacting quantum problems in one spatial dimension are the perturbative renormalization group (RG), the density matrix renormalization group (DMRG) and quantum Monte-Carlo (QMC). The RG is able to capture localization and delocalization effects for repulsive as well as for attractive interactions. However, it describes the system accurately only for small disorder strength.\cite{Giamarchi88, giam_mottglas} In this paper we make use of DMRG which tackles strong disorder and strong interactions simultaneously, and allows for a determination of the phase diagram in a broad parameter range, similar to QMC. An alternative non-perturbative method for solving correlated fermionic problems is the dynamical mean-field theory (DMFT). In combination with the geometric disorder average of the local density of states, this method allows to detect Anderson localization as well.\cite{Dobros03,  Byczuk05, hofstett_AFMdiss, Dobros10, Byczuk10} This latter approach is commonly referred to as typical medium theory. Within DMFT the self-energy is approximated to be local, which is only exact in infinite dimensions and is known to lead to qualitatively accurate results in 3D. Its real-space extension, on the other hand, treats the single-particle problem - including disorder - exactly in any spatial dimension, while the self-energy becomes site-dependent.\cite{Dobrosavljevic1998, Helmes08, snoek_trap} For this reason R-DMFT is expected to be superior to single-site DMFT in low dimensions. Here we present a quantitative analysis by a comparison of R-DMFT and DMRG results in 1D.

\subsection{DMRG}
Our implementation of DMRG \cite{White92, White93, Schollwoeck05} is based on a matrix product state variational formulation.\cite{McCulloch07,Schollwoeck11} The algorithm is formulated in the canonical ensemble (fixed number of particles $N$) and at zero temperature. It makes use of the $SU(2)$ spin symmetry of the Hamiltonian to significantly reduce the computational effort. The states are multiplet representations in the group theoretical sense, such that the actual number of different states is much larger than in a calculation with Abelian symmetries only.
We use system sizes of up to 128 sites and keep up to 600 states for the ground state calculations.

When strong disorder is present, the algorithm has the tendency of getting trapped in excited (metastable) states with a slightly higher energy than the ground state. To solve this problem we apply nonlocal changes to the wave function by adding and removing a delocalized particle in a way that minimizes the energy. In most cases, this overcomes the barrier between the metastable states and the ground state such that subsequent DMRG sweeps will converge to the true ground state. We validate this convergence by starting DMRG from different initial states chosen from either the exactly solvable case with no interaction, the homogeneous case with no disorder or the classical case with no hopping.

The single particle spectral function can be defined as
\begin{eqnarray}
\rho(\omega) &=& -\frac{1}{\pi}  \langle \text{gs}| c_i \frac{1}{\omega+i\eta+E-\mathcal{H}} c^{\dag}_i |\text{gs}\rangle \\
&& -\frac{1}{\pi} \langle \text{gs}| c^{\dag}_i \frac{1}{\omega+i\eta-E+\mathcal{H}} c_i |\text{gs}\rangle,
\end{eqnarray}
where $\ket{\text{gs}}$ denotes the ground state, $E$ is the energy of the ground state and $\eta$ the broadening constant. The calculation is done with a correction vector method. We keep up to 800 states and use a broadening constant of $\eta=0.2$. This rather large broadening saves computational time and can be subsequently improved by performing a deconvolution of the spectral function.\cite{Kleine10} 

\subsection{R-DMFT}
R-DMFT is the real-space extension of DMFT which incorporates site-dependency of the self-energy.\cite{Georges96, Dobrosavljevic1998, snoek_trap} Besides the description of the Mott-Hubbard metal-insulator transition and magnetic ordering it is capable of treating spatial inhomogeneities such as disorder. Each lattice site is mapped onto a single-impurity \mbox{Anderson Hamiltonian $\mathcal{H}_{\text{A}}$} \cite{Georges96}
\begin{eqnarray}
\mathcal{H}_{\text{A}}&=&\sum_{l\sigma}\epsilon_{l\sigma} a^\dagger_{l\sigma}a^{\phantom{a}}_{l\sigma}+\sum_{l\sigma}V_{l\sigma} \big(a^\dagger_{l\sigma} c^{\phantom{a}}_{0\sigma} + \text{h.c.}\big) \nonumber\\ &&- \mu\sum_\sigma c^\dagger_{0\sigma}c^{\phantom{a}}_{0\sigma} + Un_{0\uparrow}n_{0\downarrow}\,,
\end{eqnarray}
where $a^\dagger_{l\sigma} (c^\dagger_{0\sigma})$ and $a_{l\sigma} (c_{0\sigma})$ are fermionic creation and annihilation operators in the bath (on the impurity) and $\sigma$ represents the spin index. The parameters $\epsilon_{l\sigma}$ and $V_{l\sigma}$ determine the hybridization function
\begin{equation}\label{eq:hybrid}
\Delta_{i\sigma}(\omega)=\sum_{l}V_{il\sigma}^2\delta(\omega-\epsilon_{il\sigma})\,,
\end{equation}
which depends on the site index $i$ and is determined self-consistently.

For a given set of arbitrary hybridization functions, the solution of these effective quantum impurity problems is provided by the Numerical Renormalization Group (NRG) for $T=0$ and leads to a set of one- and two-particle on-site Green's functions, $G_{ii\sigma}(i\omega_n)$ and $F_{ii\sigma}(i\omega_n)$ respectively. These determine the self-energy matrix in real-space representation \cite{Bulla}
\begin{equation}
(\mathbf{\Sigma}_\sigma)_{ij}=\Sigma_{ii\sigma}\delta_{ij}=U\frac{F_{ii\sigma}}{G_{ii\sigma}}\,.
\end{equation}
Due to the lattice Dyson equation, the interacting lattice Green's function is given by
\begin{equation}\label{eq:int_gr}
 \mathbf{G}_{\sigma}(i\omega_n)^{-1}=\mathbf{G}^0_{\sigma}(i\omega_n)^{-1} - \mathbf{\Sigma}_\sigma(i\omega_n)\,,
\end{equation}
where $\omega_n$ are Matsubara frequencies. The noninteracting Green's function $\mathbf{G}^0_\sigma(i\omega_n)$ in real-space representation is given by
\begin{equation}
 \mathbf{G}^0_\sigma(i\omega_n)^{-1}=(\mu_{\sigma}+i\omega_n)\unity-\mathbf{J}-\mathbf{V}\,,
\end{equation}
where $\unity$ is the unity matrix, $\mathbf{J}$ is the matrix of hopping amplitudes, and $\mathbf{V}=\varepsilon_{i}\delta_{ij}$ represents the matrix of disordered on-site potentials.
Inverting Eq.~\ref{eq:int_gr} yields the interacting local Green's function
\begin{equation}\label{eq:gr_onsite}
G_{ii\sigma}(i\omega_n)=[\mu_{\sigma}+i\omega_n-\varepsilon_{i}-\Sigma_{ii\sigma}(i\omega_n)-\Delta_{i\sigma}(i\omega_n)]^{-1}\,.
\end{equation}
A set of new on-site hybridization functions $\Delta_{i\sigma}(i\omega_n)$ can be extracted from the diagonal elements in Eq.~\ref{eq:gr_onsite}. These constitute a set of new impurity problems to be solved by NRG, which closes the self-consistency loop.

\section{Results}\label{sec:results}

\subsection{Local density of states}
In order to describe the transition from delocalized to localized states it is useful to characterize the spectral properties of the system by the local density of states (LDOS), which measures the local amplitude of the wave function at a given site $i$
\begin{equation}
\rho_{i\sigma}(\omega)=-\frac{1}{\pi}\operatorname{Im} G_{ii\sigma}(\omega)\,.
\end{equation}
The calculation of the on-site Green's function $G_{ii\sigma}(\omega)$ is based on the local spectrum for a particular disorder realization $\{\varepsilon_1, \varepsilon_2,\dots,\varepsilon_L\}$. In this work we focus on the paramagnetic solution and drop the spin index $\sigma$ for readability. To gain realization-independent information, arithmetic and geometric averaging over the spectral functions is performed 
\begin{eqnarray}
\rho_{\text{a}}(\omega)&=&\frac{1}{NL}\sum_j^N\sum_i^L\rho_{i}(\omega,\{\varepsilon_1, ... ,\varepsilon_L\}_j)\\
\rho_{\text{g}}(\omega)&=&\frac{1}{N}\sum_j^N\exp\bigg(\frac{1}{L}\sum_i^L \ln\bigg[\rho_{i}(\omega,\{\varepsilon_1, ... ,\varepsilon_L\}_j)\bigg]\bigg)\nonumber\\
&&
\end{eqnarray}
given $N$ disorder realizations and $L$ sites. In this work averages over $50-100$ configurations were performed for each $(U,D)$ parameter set. The geometric average is a good approximation of the typical value of the probability distribution function for the LDOS.\cite{Dobros03} Thus, the geometrically averaged spectral function is critical at the Anderson transition, i.e. it is finite in the delocalized regime and vanishes in the localized phase and therefore can be interpreted as an order parameter.\cite{Anderson58,Dobros03,dobr,Schubert05} The arithmetic average remains finite in the Anderson-localized regime and corresponds to the global density of states (DOS).

This classification is, however, only reliable in the thermodynamic limit $L\to\infty$. In finite systems a careful analysis of finite-size effects has to be performed. In a system of length $L$ the spectrum is discrete and the non-interacting energy level spacing in the DOS scales as $1/L$. This discrete level structure due to the finite size can be smoothened by a broadening of each level to reconstruct the DOS in the thermodynamic limit. Unfortunately,  spectral broadening with a width $\eta$, which is necessary due to the finite system size, limits the spectral resolution of our calculations. The interaction-driven metal-insulator transition can therefore only be detected when the gap exceeds $\eta$. Similarly, the effect of the disorder strength is underestimated. Disorder increases the energy level spacing in the LDOS, which leads to localization. Due to broadening this discretization is smeared out and Anderson localization is observed at a larger value of $D$ than in the thermodynamic limit. 
A further finite-size effect is the competition between the system size $L$ and the localization length $\xi$. When localization sets in with increasing disorder, the localization length is larger than the system size and even exponentially localized states contribute spectral weight at all lattice sites in finite-size systems. Accordingly, disorder-driven localization can only be identified for $L\gg\xi$. Due to these limitations, the goal of our work is to reveal the localization/delocalization trends in the system rather than to determine sharp phase boundaries. 

\begin{figure}[t]
 \centering
\includegraphics[width=1.0\linewidth]{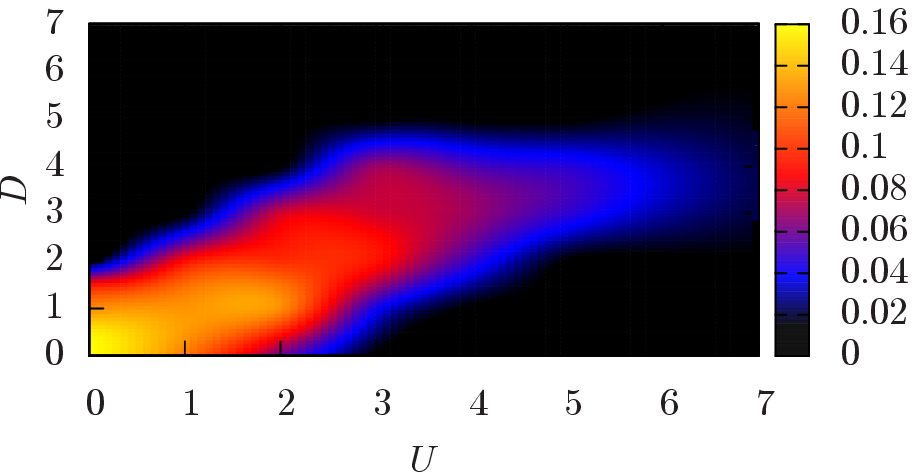}
 \centering
\includegraphics[width=1.0\linewidth]{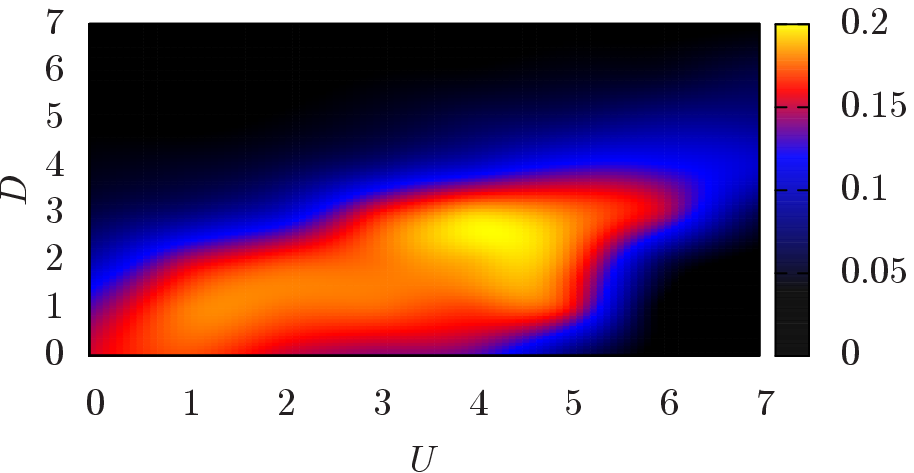}
\caption{$\rho_\text{g}(0)$ calculated by means of DMRG (upper panel) and R-DMFT (lower panel). The Mott phase is located in the lower right area and the Anderson-Mott phase in the upper left area. They are separated by a region with finite typical LDOS, which indicates a delocalization tendency.}
\label{fig:geo_rho0}
\end{figure}

To investigate a possible emergence of a metallic phase, $\rho_{\text{g}}(\omega)$ is analyzed at the Fermi level $\omega=0$ using R-DMFT with $L=128$, $\eta=0.05$ and 10-50 disorder configurations, and using DMRG with $L=64$, $\eta=0.2$ and 16 disorder configurations, where the results are additionally deconvolved. The results are presented in Fig.~\ref{fig:geo_rho0}.
\begin{figure}
\centering
\includegraphics[width=1.0\linewidth]{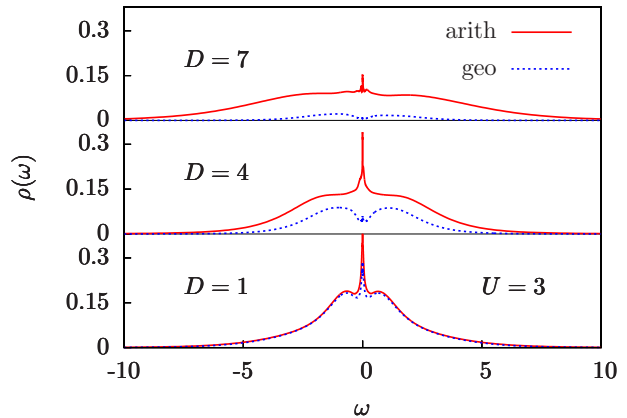}
\caption{Comparison between geometrically (dashed line) and arithmetically averaged (solid line) LDOS for $U=3$ calculated within R-DMFT. With increasing disorder strength the geometrically averaged density of states gradually vanishes, starting from the band edges.}
\label{fig:dos_U3_D1-7}
\end{figure}
\begin{figure}
\includegraphics[width=1.0\linewidth]{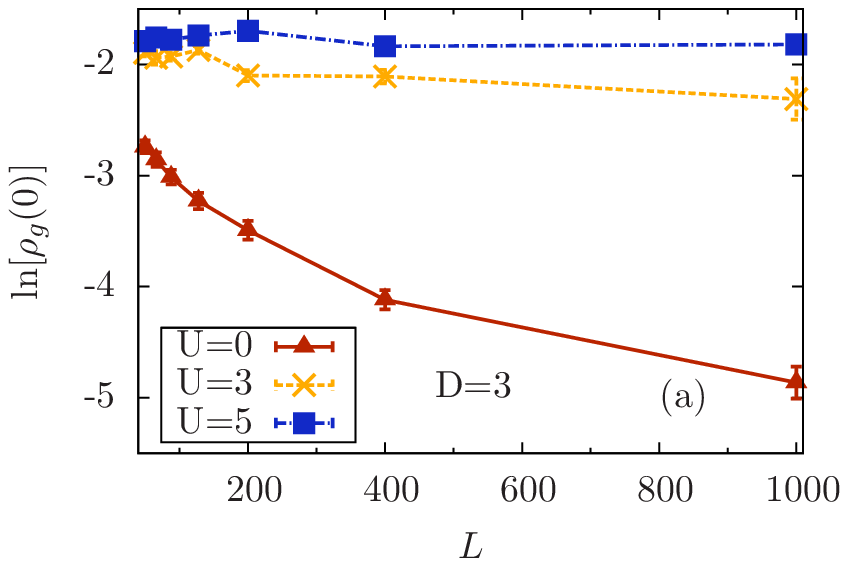}
\includegraphics[width=1.0\linewidth]{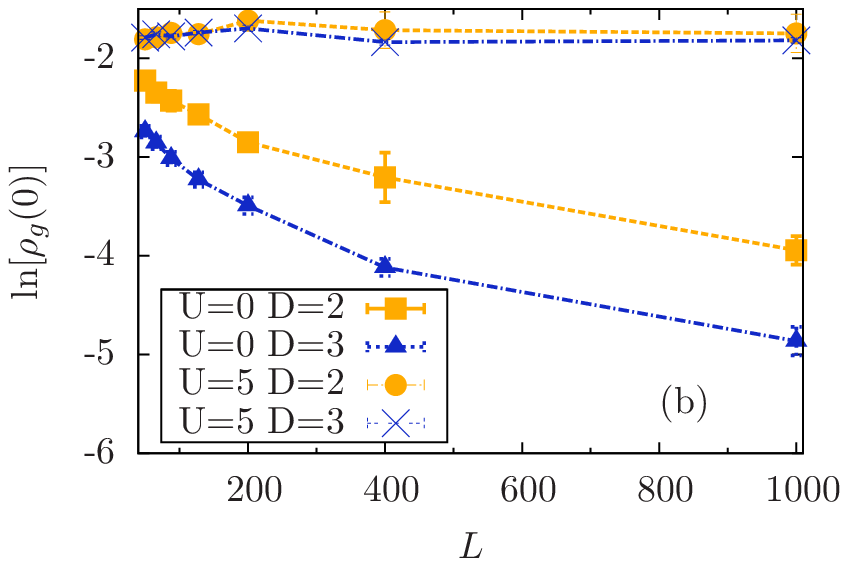}
\caption{Finite-size scaling for the geometrically averaged LDOS at the Fermi-level calculated within R-DMFT. The results are averaged over 10-50 configurations. (a) The decay with system size at $D=3$ and $U=0$ indicates Anderson localization while for $D=3$ and $U=3\,(5)$ the vanishing size dependence denotes delocalization. (b) Similarly, states which are Anderson-localized at $U=0$ become delocalized for $U=5$. Finite interaction shifts $\rho_\text{g}(0)$ to higher values and reduces the dependence on the system size.} 
\label{fig:fin-size}
\end{figure}

Three different regimes are found in the system: The Anderson-Mott insulating regime at large $D$, the Mott-insulator at large $U$ and a delocalization (metallic) tendency for intermediate disorder and interaction strength.

(i) \textbf{Anderson-Mott Insulator} (AMI): As the spectrum of the localized system consists of a dense distribution of poles,\cite{Thouless1974} the geometrically averaged LDOS vanishes at all frequencies while the arithmetical average remains finite. The formation of AMI proceeds from the edges of the band towards the band center with increasing disorder strength (see Fig.~\ref{fig:dos_U3_D1-7}). As soon as the geometrically averaged LDOS $\rho_{g}(\omega)$ vanishes at the Fermi level $\omega=0$ the system becomes fully localized. We first analyze our results along the $D$ axis at $U=0$. Exact calculations predict the non-interacting system to undergo an Anderson metal-insulator transition in the thermodynamic limit $L\to\infty$ at any finite disorder strength.\cite{Lieb68} However, we obtain clear signatures of localization at $U=0$ only for $D\gtrsim2$ (DMRG) or $D\gtrsim3$ (R-DMFT). This can be traced back to the finite size of the investigated system. The localization length in this case is larger than the system size, such that the states remain quasi-extended  and are only localized for very large system sizes.\cite{lagendijk} A careful analysis of the scaling behavior with system size was performed in R-DMFT for $U=0$ and $D=3$ and is displayed in Fig.~\ref{fig:fin-size}a. The decay of $\rho_{g}(0)$ with increasing system size confirms localization (see a detailed discussion of the finite-size scaling below). 

(ii) \textbf{Mott Insulator} (MI): Here for $D=0$ and $U>0$ the LDOS $\rho(\omega)$ vanishes at the Fermi level $\omega=0$ due to the appearance of a Mott gap. In the deconvolved DMRG results $\rho$ decreases monotonously with increasing interaction and a MI appears for $U\gtrsim3$. Within R-DMFT the Luttinger theorem ensures pinning of the spectral weight at the Fermi level.\cite{Luttinger} Thus, $\rho_{g}(\omega=0)$ remains finite and constant up to $U=4$ and the gap is first formed at $U\gtrsim5$ (see Fig.~\ref{fig:geo_rho0}). DMRG as well as R-DMFT results for finite system size disagree with exact calculations for a homogeneous system in thermodynamic limit, where a Mott insulator is predicted to appear at any finite interaction strength.\cite{Lieb03} In R-DMFT, as in every other DMFT extension, the finite metallic phase  in 1D is due to the approximation of a local self-energy, which becomes exact only in infinite dimensions. In DMRG this discrepancy in the detection of the Mott gap is a consequence of the unavoidable spectral broadening $\eta$. Due to this broadening the Mott-gap is smeared out and the resulting critical $U$ represents an upper bound at which the Mott-insulator is formed. In DMRG, the broadening effects are partially corrected by applying a deconvolution to the spectra. A careful analysis of the finite-size effects in the gap formation is presented in section \ref{sec:charge_gap}. 

(iii) \textbf{Delocalization}: For low to intermediate disorder and interaction, the geometrically averaged spectral density $\rho_\text{g}(\omega)$ is finite at the Fermi level $\omega=0$. The non-black area in both phase diagrams in Fig.~\ref{fig:geo_rho0} indicates this delocalization regime. Due to the deconvolution procedure within DMRG the spectral weights are lower than those obtained by R-DMFT. 

In the absence of disorder and interaction ($U=D=0$) the system is metallic and satisfies Luttinger's theorem\cite{lutt} in agreement with our findings. Additionally, a delocalization region is observed for intermediate interaction and disorder strengths. The latter, however, may in principle be due to overestimated spectral weights in finite systems for localization length $\xi>L$ and finite spectral broadening.

To decide whether true delocalization is observed, a careful finite-size scaling analysis of the R-DMFT results was performed for the intermediate $U$ and $D$ regimes. The scaling behavior of $\rho_\text{g}(0)$ characterizes the phase of the system: in the non-interacting case if the states are Anderson-localized  and the localization length $\xi\ll L$, the geometrically averaged LDOS scales as $\exp[-L/\xi]$ for periodic boundary conditions. If $\xi> L$ the decay of $\rho_\text{g}(0)$ with system size is algebraic.\cite{Janssen98} In systems with finite interactions the functional dependence of $\rho_\text{g}(0)$ on system size is not known, however, the geometrically averaged spectral weight is still expected to decay to zero with increasing system size for localized states and to remain finite for delocalized ones in the limit $L\to\infty$.

In Fig.~\ref{fig:fin-size}a we observe the delocalizing effect of interactions. For $D=3$ and $U=0$ the geometrically averaged LDOS decays with increasing system size, which indicates the Anderson localized phase. Thus, the final spectral weight visible in the
phase diagram in Fig.~\ref{fig:geo_rho0} (lower panel) for this parameter set is a finite-size effect for the $L=128$ lattice.
Upon increasing the interaction strength to $U=3$ and $U=5$ no clear localization signature could be observed anymore within accessible system sizes.

In Fig.~\ref{fig:fin-size}b the effects of increasing disorder strength and interactions on the localization properties of the system are compared. A geometric average which decays with increasing system size $L$ for $U=0$ and $D=2$ and $3$ indicates the Anderson localized phase. Again, the delocalization trend observed in Fig.~\ref{fig:geo_rho0} (lower panel) for this parameter set turns out to be a finite-size effect for the $L=128$ lattice. The redistribution of spectral weight due to increased disorder strength leads to a well pronounced shift of $\rho_\text{g}(0)$ towards lower values for $D=3$ compared to $D=2$ in the non-interacting case. Strong interaction $U=5$ shifts the geometric average upwards and the dependence on the system sizes studied here vanishes which indicates delocalization. 

\subsection{Charge gap}\label{sec:charge_gap}
In the homogeneous 1D Hubbard model at half-filling a metallic phase exists only at $U=0$. At any finite interaction strength a charge gap in the density of states is predicted to appear and the system becomes Mott insulating. Thus, starting in the MI phase, the additional influence of disorder can be detected via a vanishing gap. The charge gap is calculated by means of DMRG as $ G = [E(N+1)+E(N-1)-2E(N)]$, where $E(N)$ is the ground state energy for $N$ particles. It is important to note that the computation of the energy $E(N)$ is not based on the spectrum and is therefore not affected by the artificial broadening $\eta$. Thus, the calculated charge gap is exact for a given system size. We use system sizes up to $L=128$ sites and average over $16$ disorder realizations. 

\begin{figure}[t]
\centering
\includegraphics[width=1.0\linewidth]{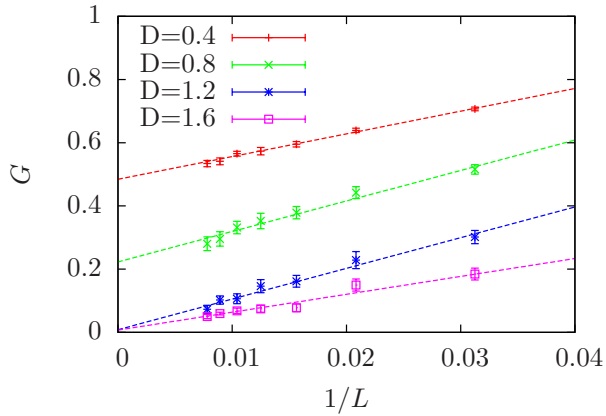}
\caption{Charge gap versus inverse system size for $U=3$ calculated within DMRG. For disorder strength $D\gtrsim1.5$ the disappearance of the gap in the thermodynamic limit $1/L\to 0$ indicates the transition from the gapped Mott insulating to a gapless delocalized phase.}
\label{fig:gap_infty_scaling}
\end{figure}

The finite-size scaling for the charge gap at $U=3$ and various disorder strengths is presented in Fig.~\ref{fig:gap_infty_scaling}. In the thermodynamic limit $1/L\to0$ the charge gap closes with increasing disorder and vanishes at $D\approx 1.5$. The scaling of the charge gap with system size was fitted by the function
$G(L) = G_\infty+\frac{a}{L}$. This finite-size analysis confirms the delocalization in Fig.~\ref{fig:geo_rho0} (upper panel) for $U=3$ and $D\gtrsim1.5$. In contrast, the finite spectral weight at smaller disorder strength was found to be a finite-size effect. 

\begin{figure}[t]
\centering
\includegraphics[width=1.\linewidth]{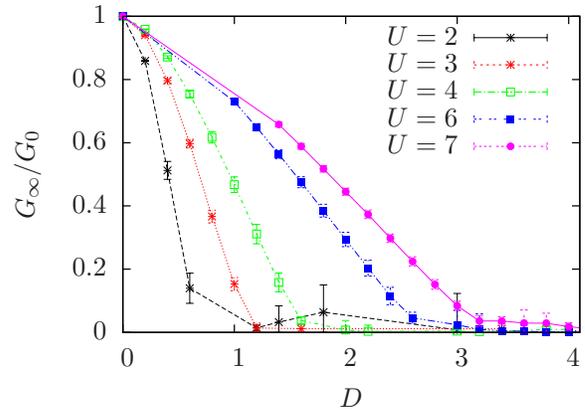}
\caption{Scaling of the gap normalized to the non-disordered values $G_0$ in the thermodynamic limit, calculated within DMRG. The gap effectively vanishes for disorder strength $D\approx U/2$. The error bars correspond to the statistical errors of the fitting routine.}
\label{fig:gap_infty_new}
\end{figure}

The extrapolated gaps for various disorder and interaction strengths normalized with respect to the extrapolated gaps of the corresponding homogeneous systems  are presented in Fig. \ref{fig:gap_infty_new} for different interaction strength. The indicated error bars are the asymptotic standard errors of the fitting routine. For intermediate interaction strength the critical disorder needed to destroy the gap was found to be $\approx U/2$. This tendency becomes less clear for $U\leq2$ as the limited resolution of the exponentially small gap complicates our analysis. The remaining gapless region covers the delocalized as well as the Anderson-Mott localized regime. For a quantitative analysis of this region we calculate the inverse participation ratio as described in the following section.

\subsection{Inverse participation ratio}
\begin{figure}[t]
\centering
\includegraphics[width=1.0\linewidth]{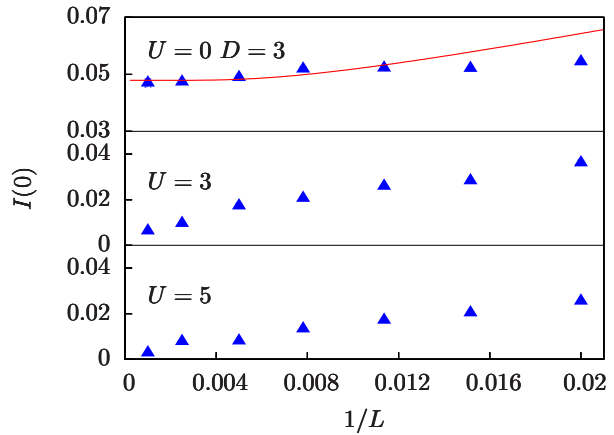}
\caption{(Color online) Finite-size scaling of the IPR, calculated within R-DMFT. The $U=0$, $D=3$ data are fitted by $I(0)=I_{\infty}(\omega)\coth[L/\xi(\omega)]$ which corresponds to the characteristic scaling in the localized regime. The decay of $I(0)$ for $D=3$ and $U=3\,(5)$ with increasing system size indicates delocalization.}
 \label{fig:ipr}
\end{figure}
The localization properties of the system can be quantified by means of the inverse participation ratio $I(\omega)$ (IPR).  This observable corresponds to the inverse of the number of sites over which a state is extended. In the non-interacting homogeneous case it is defined as  $I(\omega)=\sum_i^L|\psi_i(\omega)|^4$, where $\psi(\omega)$ is a single particle wave function. If a state at frequency $\omega$ is exponentially localized with corresponding localization length $\xi(\omega)$  then $I(\omega)$ scales as $I_{\infty}(\omega)\coth[L/\xi(\omega)]$ with the system size for periodic boundary conditions and $\lim_{L\to\infty}I(\omega)= I_{\infty}(\omega)=1/\xi(\omega)$. In the case of a purely delocalized state, i.e. for a wave function homogeneously extended over the whole system, the IPR vanishes as $1/L$ with increasing system size.

In an interacting system the IPR is defined as \cite{YunSong}
\begin{equation} \label{eq:ipr}
 I(\omega) = \frac{\sum_i^L \rho_i(\omega)^2}{(\sum_i^L \rho_i(\omega))^2} \,.
\end{equation}
Similarly to the non-interacting case, in the thermodynamic limit the delocalized states lead to a vanishing IPR, while in the localized system the IPR remains finite for $L\to\infty$. Thus, the scaling behavior with system size gives insight into localization or delocalization driven by disorder or interaction.
In Fig.~\ref{fig:ipr}  the IPR at $\omega=0$ is plotted as a function of system size. As expected, for $U=0$ and $D=3$ the system is Anderson localized and the IPR saturates for $L\to\infty$. Moreover, fitting the data by $I(0)=I_{\infty}\coth[L/\xi(\omega)]$ shows that the scaling behavior also holds for open boundary conditions at larger system sizes. For finite interaction strength $U=3$ and $U=5$, however,  $I(0)$ decays with system size (note the differing scales on the y axis). The vanishing IPR highlights the delocalizing effect of the repulsive interaction. These results confirm our findings in Fig.~\ref{fig:fin-size}.

\section{Summary}\label{sec:summary}
We have studied the occurrence of delocalization, Anderson- and Mott-insulating phases for interacting disordered spinful fermions on a 1D lattice at commensurate filling. The phases were characterized by means of the local density of states, the scaling of the inverse participation ratio and the charge gap. Both numerical techniques used here, DMRG and R-DMFT, agree in their predictions for parameter regimes well within the Anderson and Mott phases. However, R-DMFT was found to be more affected by finite-size effects than DMRG, which is reflected in a larger delocalization trend for fixed lattice size. As our main result, we present the first non-perturbative calculations indicating a delocalized phase of spinful fermions in 1D due to the interplay of disorder and interaction of intermediate strength, similar to the bosonic case. This trend, observed in DMRG as well as R-DMFT, persisted  in finite-size scaling within both methods. 
 
\begin{acknowledgments}
The authors acknowledge helpful discussions with A. Kleine, D. Semmler and I. Titvinidze. This work was supported by the DFG research unit FOR 812. WH acknowledges the hospitality of the Aspen Center of Physics during the final stage of this work, supported by the National Science Foundation under Grant No. 1066293.
\end{acknowledgments}	

%

\end{document}